\documentclass[aps,pra,twocolumn,nopacs,superscriptaddress,groupedaddress]{revtex4}

\usepackage{amsmath}
\usepackage[dvips]{graphicx}
\usepackage{epsfig}
\usepackage{tabularx}
\usepackage{color}
\usepackage[pdfpagemode=UseNone,colorlinks=true,linkcolor=blue,citecolor=blue]{hyperref}

\usepackage{soul}
\usepackage{gensymb}

\renewcommand\Re{\operatorname{Re}}
\renewcommand\Im{\operatorname{Im}}

\begin{document}

\title[]{Feedback cooling scheme for an optically levitated oscillator with controlled cross-talk}

\author{J.M.H. Gosling}%
\email{jonathan.gosling.14@ucl.ac.uk}
\affiliation{Department of Physics and Astronomy, University College London, Gower Street, London WC1E 6BT, United Kingdom}%

\author{A. Pontin}
\email{a.pontin@ucl.ac.uk}
\affiliation{CNR-INO, largo Enrico Fermi 6, I-50125 Firenze, Italy}

\author{F. Alder}
\affiliation{Department of Physics and Astronomy, University College London, Gower Street, London WC1E 6BT, United Kingdom}

\author{M. Rademacher}
\affiliation{Department of Physics and Astronomy, University College London, Gower Street, London WC1E 6BT, United Kingdom}

\author{T. S. Monteiro}%
\affiliation{Department of Physics and Astronomy, University College London, Gower Street, London WC1E 6BT, United Kingdom}%

\author{P. F. Barker}
\email{p.barker@ucl.ac.uk}
\affiliation{Department of Physics and Astronomy, University College London, Gower Street, London WC1E 6BT, United Kingdom}%

\begin{abstract}
Levitated optical mechanical systems have demonstrated excellent force and impulse sensitivity and are currently being developed for the creation of non-classical states of motion in these new quantum systems. An important requirement in the design of these systems is the ability to independently control and cool all three translational degrees of freedom. Here we describe the design and implementation of a stable and robust 3D velocity feedback cooling scheme with particular emphasis on creating minimal cross-talk between the independent oscillatory modes when cooling. 
\end{abstract}

\maketitle

\emph{Introduction} - Methods for cooling optically levitated nano-oscillators, that are well decoupled from the environment in high vacuum, are under active development for ultrasensitive force sensors that could operate at the quantum limit~\cite{liang2023yoctonewton,ranjit2016zeptonewton} and also for the creation of macroscopic non-classical states of motion~\cite{Magrini2021real,Novotny2021quantum,piotrowski2023simultaneous,delic2020cooling,dania2024high}.  

The development of efficient methods for cooling these oscillators are an important requirement, either to avoid the non-linear motion in the optical trap at high temperatures, or to reach the quantum ground state within the confining optical potential.  Already some cooling schemes are capable of reaching the ground state via both feedback damping and cavity cooling methods ~\cite{tebbenjohanns2019cold,tebbenjohanns2020motional,kamba2021recoil,dania2021optical,kamba2022optical,vijayan2023scalable,liska2023cold}. 
Of particular importance in designing these schemes is the ability to cool each degree of freedom as independently as possible, since inevitable cross-coupling induced in the cooling process can reduce the efficiency and eventually limit the lowest temperatures that can be achieved. This is especially important when the modes are not well separated in frequency or for directional force searches~\cite{Gosling2024Sensing}, where induced cross-talk can mask the signature of a directional force. Feedback methods are often desirable for these applications, since each oscillation mode can be in principle cooled independently. While both cold damping and parametric feedback cooling has been demonstrated, cold damping is arguably more effective than parametric cooling~\cite{penny2021performance} and this method has been used to cool both the three translational and rotational degrees of freedom in an optical trap~\cite{iwasaki2019electric,kamba2023nanoscale}. 

Although the three directional components of a charged levitated oscillator have been successfully cooled via the Coulomb force~\cite{Kremer2024All}, the often complicated fields created by electrodes within the small trapping region can create cross-talk between the mechanical modes of the nano-oscillator. Here we present a new method of implementing cold damping which can significantly reduce the degree of cross-talk by an order of magnitude between the mechanical modes allowing us to cool in 3D to the 100~$\mu$K range limited by only by detection efficiency and vacuum.

\begin{figure}
\includegraphics[width=0.8\columnwidth]{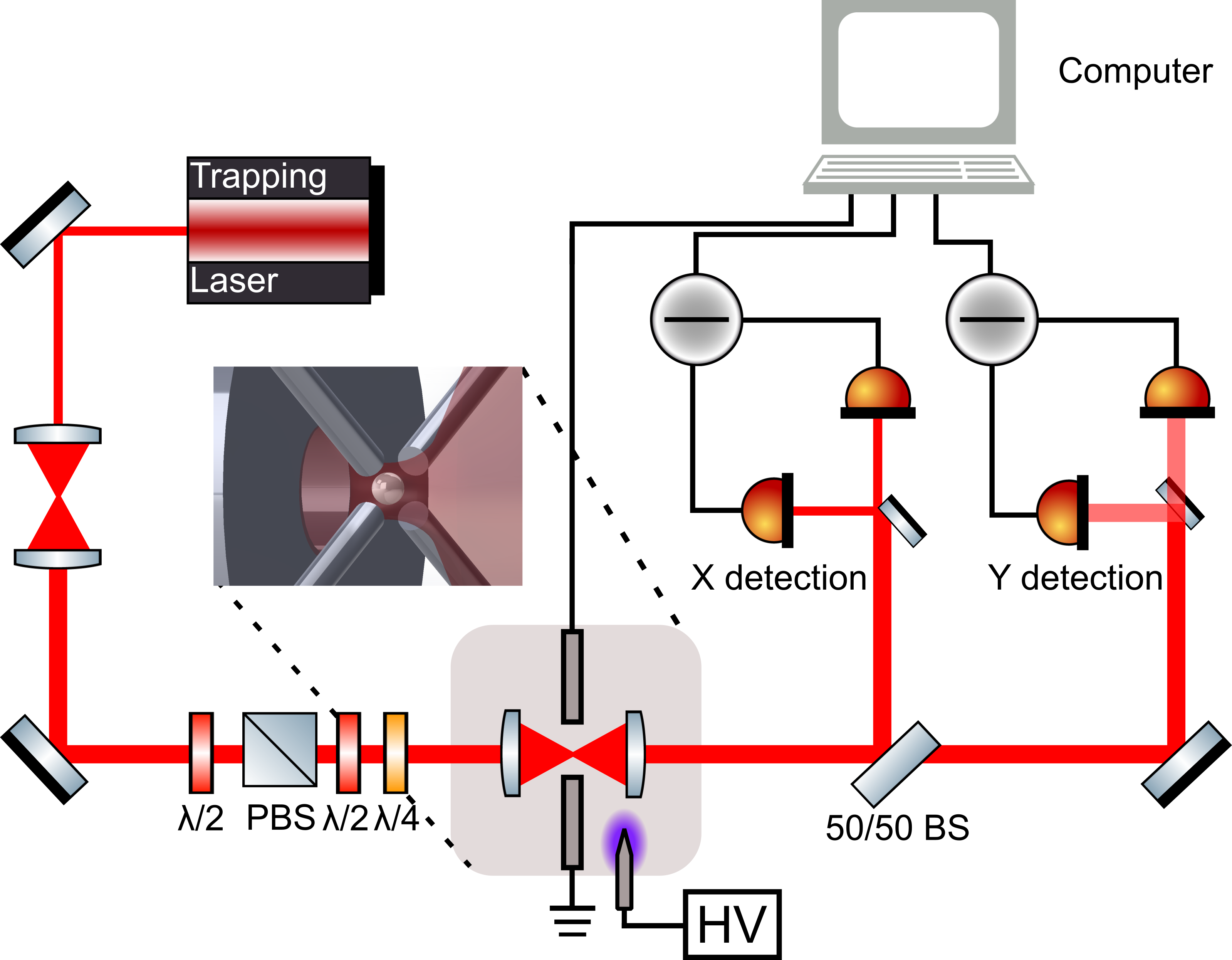}
\centering
\caption{\label{fig:figLayout}Schematic view of experimental layout with inset of the electrode configuration. The optical tweezer was formed by the tight focusing of the laser beam by a high numerical aperature (NA=0.77) aspheric lens. The scattered light from the nanoparticle is collected by another aspheric lens (NA=0.55).}
\end{figure}

\emph{An optically levitated oscillator with 3D cooling} - A schematic diagram of our setup used for levitation by optical tweezers is shown Fig.~\ref{fig:figLayout}.  Here an optical tweezer is formed by tightly focusing a Gaussian beam laser beam at a wavelegth of $\lambda=1064$\,nm with a high numerical aperture aspheric lens (Lightpath 355330, NA=0.77). Another aspheric lens (Thorlabs A230TM-B, NA=0.55) was used to collect the scattered light from the nanoparticle to determine its time dependent displacement. Silica nanospheres of radius $\approx80\,$nm were loaded into the trap at atmospheric pressures via a nebuliser, whilst the trap laser light was linearly polarised. Electrodes surround the nanoparticle in the trap in order to provide an electric field that has a component along all three spatial dimensions. This allows us to provide a Coulomb force on the charged nanoparticle in each degree of freedom of the translational motion. The nanoparticles loaded into the trap are naturally charged with one to a few elementary charges. Detection of the motion in the $x$ and $y$ direction was achieved using a split detection method where the scattered light was divided into two directions by orthogonally orientated D-mirrors~\cite{Li20213D}. The forward-propagating light is split evenly and sent to two photodiodes. The difference between these two signals was taken which provides a current which is proportional  to the displacement of the particle in one degree of freedom.  

After the nanoparticle is trapped, the pressure in the vacuum chamber is evacuated to approximately 5 mbar where the size of the nanoparticle can be characterised by thermal forces at equilibrium~\cite{Hebestreit2018Calibration}. The investigations into the cross-talk were carried out at 0.5 mbar. This pressure was chosen as it was low enough that the effects of the feedback could be observed without the effects of trap non-linearities seen at lower pressures.

The voltage signal used to cool the nanoparticle's motion was a feedback signal derived from the position measurement obtained from the forward scattered light as described above. At low pressures, the nanoparticle's motion can be approximated as sinusoidal, $q(t)=A \cos(\omega_0 t)$ and the velocity as $\Dot{q}(t)=-\omega_0 A \sin(\omega_0 t)=A\omega_0 \cos(\omega_0 t -\pi/2) $. 

To experimentally access the velocity signal, a 90-degree phase shift of the position measurement is required. This can be achieved by using the group delay from a bandpass filter with central frequency at the oscillation frequency which also filters out undesirable noise at other frequencies.
A field programmable gate array (RedPitaya STEMlab 125-14) running the PyRPL software package~\cite{pyrpl} was used to generate the bandpass filter. This software allowed control over the central frequency, filter bandwidth, the gain and the phase of the output signal. Control of the phase was also needed to accommodate for additional phase delays introduced by the electronics.

\emph{Cold damping of a charged nanoparticle} - 

We consider a cold damping scheme on a single charged nanoparticle levitated in an optical tweezer. The dynamics of the centre-of-mass motion was determined by thermal noise from gas collisions as well as the feedback force with its associated technical and fundamental noise contributions and eventually recoil heating. The feedback force we consider exploits the Coulomb interaction so that $\textbf{F}_{fb}=q\,\textbf{E}_{tot}$ where $q$ is the charge of the particle and the total electric field, $\textbf{E}_{tot}$, which depends on the specific geometry chosen for the electrodes. We begin by illustrating the scheme for a 1D oscillator~\cite{Li2011Millikelvin, penny2021performance,poggio2007feedback}, before moving to a full 3D oscillator where additional considerations must be taken. We consider an equation of motion for the levitated particle as
\begin{equation}\label{eq1:initial}
  \ddot{u}+\gamma\,\dot{u}+\omega_u^2 u=(f_{\text{th}}+q E)/m 
\end{equation}

\noindent where $u$ is the displacement in 1D, $\gamma$ is the gas damping rate~\cite{Epstein1924on,Cavalleri2010gas} arising from collisions with background gas, $\omega_u$ the oscillator frequency, and $q$ and $m$ are the charge and mass of the particle respectively. The thermal force noise is given by $f_{\text{th}}$. The electric field $E$ used to apply force to the nanoparticle can be written as $E= T V_{fb}$, where $T$ is the inverse of an effective distance which converts an applied voltage to electric field and $V_{fb}$ is the electric signal for the feedback.  To obtain a signal that is proportional to displacement, one has to perform a measurement which will have a transduction coefficient $c_u$ (in V/m) and a sensitivity limit $n_u$, given primarily by the intrinsic imprecision noise of the measurement. Consequently, the measured position can be written as $u_{mes}=c_u u+n_u$. This signal then undergoes processing to optimise the parameters for the feedback which we encapsulate in the transfer function $H(\omega)$ that typically includes the effects of amplification and delay. One must also consider any additive voltage noise $n_{el}$ associated with this signal processing. Here the electric field at the oscillator position can be written as $E=T [ n_{el}+H \ast (c_u u +n_u)]$ where $ \ast $ represents a convolution product. By considering the Fourier transform of the acceleration and velocity terms, FT, ($\textbf{FT}(\ddot{u}(t))=-\omega^2u(\omega)$ and $(\textbf{FT}(\dot{u}(t))=-i\omega u(\omega)$ and substituting in the definition of the applied electric field, Eq.~\ref{eq1:initial} can be expressed as:
\begin{equation}\label{eq:6.2}
\begin{split}
    &u(\omega)(\omega^2_u -\omega^2 -i\omega\gamma_{CM}) \\
    &=\frac{1}{m}(f_{th}(\omega)+qT(n_{el}+H(\omega)[c_u u(\omega) +n_u]))
\end{split}
\end{equation}
Collecting the position and force terms gives:
\begin{equation}\label{eq:6.3}
\begin{split}
   & u(\omega)(\omega^2_u -\omega^2 -i\omega\gamma_{CM}-\frac{1}{m}qTH(\omega)c_u) \\
   & =\frac{1}{m}(f_{th}(\omega)+qT(n_{el}+H(\omega)n_u))
\end{split}
\end{equation}         
Eq.~\ref{eq1:initial} can be described in Fourier space as:
\begin{equation}\label{eq2:fourier}
  u(\omega)= \chi_u(\omega) N_u(\omega)
\end{equation}
\noindent where $N_u(\omega)=f_{th}+qT(n_{el}+H(\omega) n_u)$ is a linear combination of all noise sources which are assumed to have a white noise spectrum. We have defined the effective susceptibility as $\chi_u(\omega)=[m(\omega_u^2-\omega^2+i\omega \gamma-g_{uu}(\omega)/m)]^{-1}$ with $g_{uu}(\omega)=q T H(\omega) c_u$.

It is useful to estimate the effective parameters for the harmonic motion. From the definition of the susceptibility we have the effective frequency $\omega_{eff}=(\omega_u^2-\Re[g_{uu}(\omega_u)/m])^{1/2} $ and an effective damping of $\gamma_{eff}=\gamma-\Im[g_{uu}(\omega_u)/(m\omega)]$. In simple terms, for cooling efficiency the objective is to minimize the resonance shift and automatically maximize the feedback damping. This is typically achieved by optimizing the transfer function $H$. 

\emph{Minimising cross-talk in 3D cooling - }
The extension of the description to the 3D case is straight forward provided that the possibility of cross coupling between the modes is carefully take into account. In a 3D space the effective distance $T$ becomes a vectorial quantity with components reflecting the electrodes geometry. For each centre-of-mass direction $u=x,y,z$, we define a vector $\textbf{T}_u=T_{ux}\, \textbf{e}_x + T_{uy}\, \textbf{e}_y + T_{uz}\, \textbf{e}_z$. The total electric field can be written as  $\textbf{E}_{tot}=\textbf{T}_x V_{fb,x}+\textbf{T}_y V_{fb,y}+\textbf{T}_z V_{fb,z}$ where $V_{fb,i}$ is the electric signal for the feedback along different directions.

The system of equations for the centre-of-mass motion can then be written in Fourier domain in the following form
\begin{equation}\label{eq3:motion2}
\begin{split}
  x&=\chi_x^0\, (g_{xx}\, x+ g_{xy}\, y+g_{xz}\, z+ N_x) \\
  y&=\chi_y^0\, (g_{yx}\, x+ g_{yy}\, y+g_{yz}\, z+ N_y)\\
  z&=\chi_z^0\, (g_{zx}\, x+ g_{zy}\, y+g_{zz}\, z+ N_z),
\end{split}
\end{equation}

\noindent where $\chi_i^0=[m(\omega_i^2-\omega^2+i\omega \gamma)]^{-1}$ is the mechanical susceptibility without feedback present with $g_{i,j}=q\,T_{j,i} H_j c_j$, representing the i-component of the effective distance along the j-axis. However, when modeling a typical tweezer experiment, Eq.~\ref{eq3:motion2} can be further simplified. Since $\omega_z\ll \omega_x,\omega_y$, cross couplings between $z$ and $x$ and $y$ can be neglected, this allows us to write the following solution for each direction
\begin{equation}\label{eq4:sol}
\begin{split}
  x&=\zeta_x\, (g_{xy}\,\chi_y N_y + N_x) \\
  y&=\zeta_y\, (g_{yx}\,\chi_x N_x + N_y)\\
  z&=\chi_z\,  N_z,
\end{split}
\end{equation}

\noindent where $\zeta_i=\chi_i (1-g_{ij} g_{ji} \chi_i \chi_j)^{-1}$. It is clear from Eq.~\ref{eq3:motion2} and Eq.~\ref{eq4:sol} that the degrees of freedom can potentially become coupled depending on the value and functional form of the cross terms $g_{i,j}$. As we will see in the next section these depend on the chosen geometrical configuration for the feedback electrodes and can be made vanishing small with the appropriate choice of electrode orientation.

\begin{figure}
\centering
\includegraphics[width=1\columnwidth]{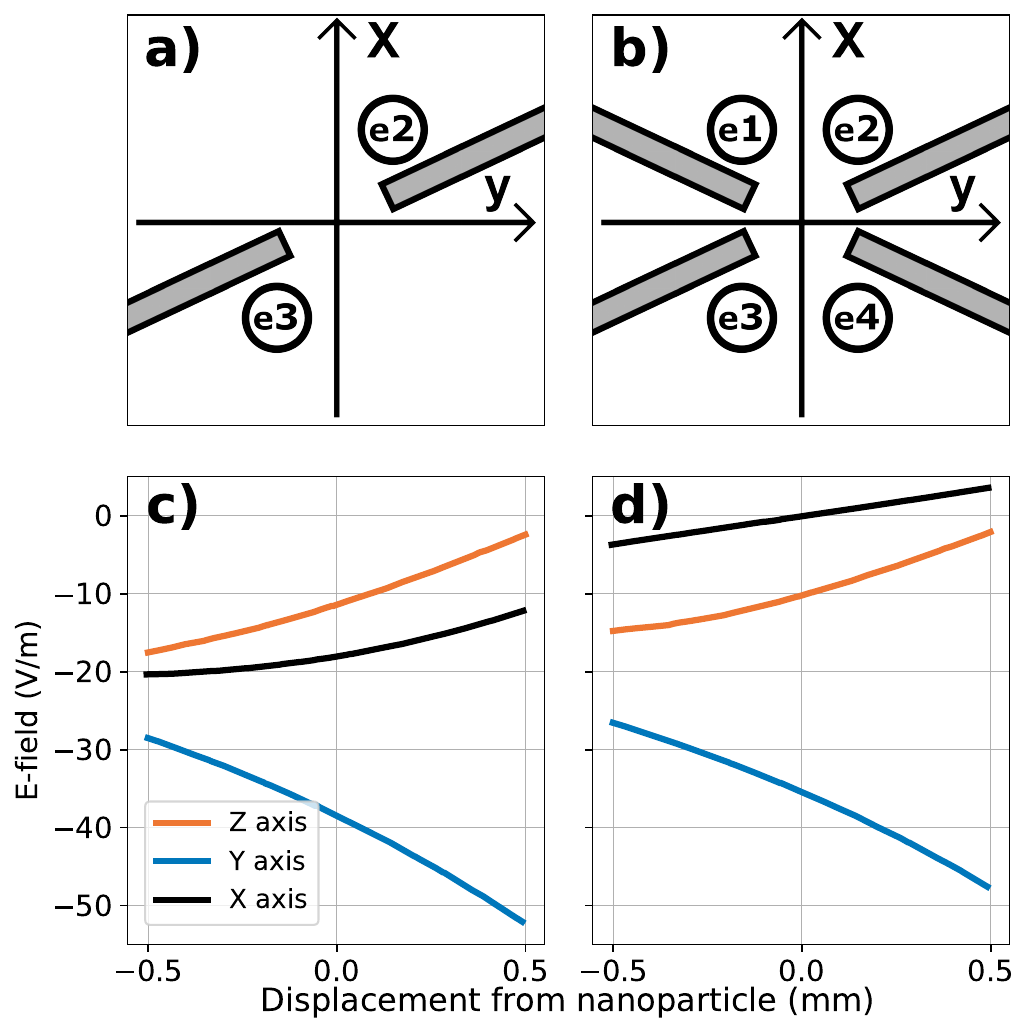}
\centering
\caption{\label{fig:figureVD1} Schematic overview of the electrodes geometries considered and finite element simulations for the resulting electric fields. a) one drive electrode configuration. b) three drive electrodes geometry. In both cases, the electrodes are cylindrical and lie in a plane parallel but slightly shifted from the $x-y$ plane. The origin of the reference frame is placed at the nominal particle mean position. The tweezer field is considered to propagate in the $+z$ direction. c) electric field components along their respective axis when an electrostatic potential of $1$\,V is applied to electrode 2 and 3 is grounded. d) As panel c) but with a $1$\,V potential applied to electrodes 4 and 2 while the other two electrodes are grounded. 
Note the electric field for the plot on the right is halved to ensure that the plot have the same electric field range.}
\end{figure}

\emph{Electrode geometry} - We consider two electrode geometries. The first, which is the most common in the literature \cite{ricci2019accurate,iwasaki2019electric}, consists of two opposing coaxial electrodes oriented at an angle with the $x$ axis. To achieve 3D velocity damping, a single drive electrode is paired with a corresponding ground electrode, as shown in Fig.~\ref{fig:figureVD1}\,a). The electrodes lie in the $xy$ plane and the angle ensures there is a component of the electric field that acts in both the $x$ and $y$ directions. This electrode is also used to cool the $z$ motion by grounding the metal holder of the optics used to trap the nanoparticle so that the field lines bend in the direction of propagation.

The second configuration, consists of four electrodes, as show in Fig.~\ref{fig:figureVD1}\,b). Here, the feedback signal to cool the $x$ and $y$ motion is applied to electrode 4 and 1 respectively, while electrode 2 is driven with the sum of the two signals.

The electric field obtained at the nanoparticle position can be estimated with numerical simulations based on the finite element method.  The results are shown in Fig.~\ref{fig:figureVD1} c) and d) where the electric field components are plotted along the respective axis (e.g., $E_x$ along the $x$ axis ecc.). For both configurations, an electrostatic potential of $1$\,V is assumed to be driving the relevant electrodes. The first configuration is shown in Fig.~\ref{fig:figureVD1}~c) where it is clear that a signal will generate an electric field component in all three directions.

On the other hand, Fig.~\ref{fig:figureVD1}~d) shows the 4 electrodes configuration when  $e2$ and $e4$ are driven, i.e., the electrodes for the $y$ feedback. In this case, the electric field component along $x$ vanishes. This can be understood from simple symmetry considerations. When driven with a signal $V_d$, $e2$ will generate a field $\textbf{E}_2=V_d\,(T_x,T_y,T_z)^T$ while $e4$ will give $\textbf{E}_4=V_d\,(-T_x,T_y,T_z)^T$ so that their sum will be $\textbf{E}=V_d\,(0,2 T_y,2 T_z)^T$. The same argument can be used to show that the feedback signal for $x$, applied on $e1$ and $e2$, will give a vanishing field component along the $y$ axis. It is important to note that the electric field plotted in Fig.~\ref{fig:figureVD1}~d) is halved to have the same electric field range.

The two configurations are in many respects equivalent except in those applications where induced cross-talk is greatly detrimental, as, e.g., in the case of directional force sensing~\cite{Gosling2024Sensing,Pontin2023Controlling}. 

As shown before, the cross-coupling terms are given by $g_{i,j}=q\,T_{j,i} H_j c_j$. The 4 electrodes configuration suppresses the hybridization by ensuring the coefficients $T_{j,i}$ are vanishingly small. Indeed, any deviation from zero can only emerge from an imperfect geometry and/or a shifted particle mean position. In the case of the 2 electrodes configuration the hybridization is unavoidable.

\emph{Experimental implementation} - To demonstrate this experimentally, a levitated nanoparticle was cooled using both electrode configurations. A schematic layout of the experiment was shown in Fig.~\ref{fig:figLayout}.  The transfer function $H(\omega)$ we use a
band-pass filters, in our case a digital filter generated by the field programmable gate array with transfer function

\begin{equation}\label{eq5:transferfunction}
    H(\omega)=a_0 \Gamma \frac{i \omega}{\omega^2-\omega_0^2+i \omega \Gamma}  e^{i\phi}
\end{equation}

\noindent which is equivalent to a resonant RLC filter with centre frequency $\omega_0$ and full linewidth $\Gamma$ with the addition of a gain $a_0$ and arbitrary delay set by the phase $\phi$.

 The feedback signals were generated by two FPGAs, one associated with each of the detectors. For the one drive electrode configuration, the output of the FPGA for the $x$ and $z$ motion is fed to the input of the $y$ FPGA to combine the three output signals. This signal is then amplified and is applied to one electrode. For the three drive electrode case, the signals from the two FPGAs are also summed to obtain the signal for the $e$2. The gain settings used in the one electrode case were adjusted to compensate for the stronger electric fields generated in the three electrode case to ensure that a comparable cooling was achieved in both experimental runs.

\begin{figure}
\includegraphics[width=1\columnwidth]{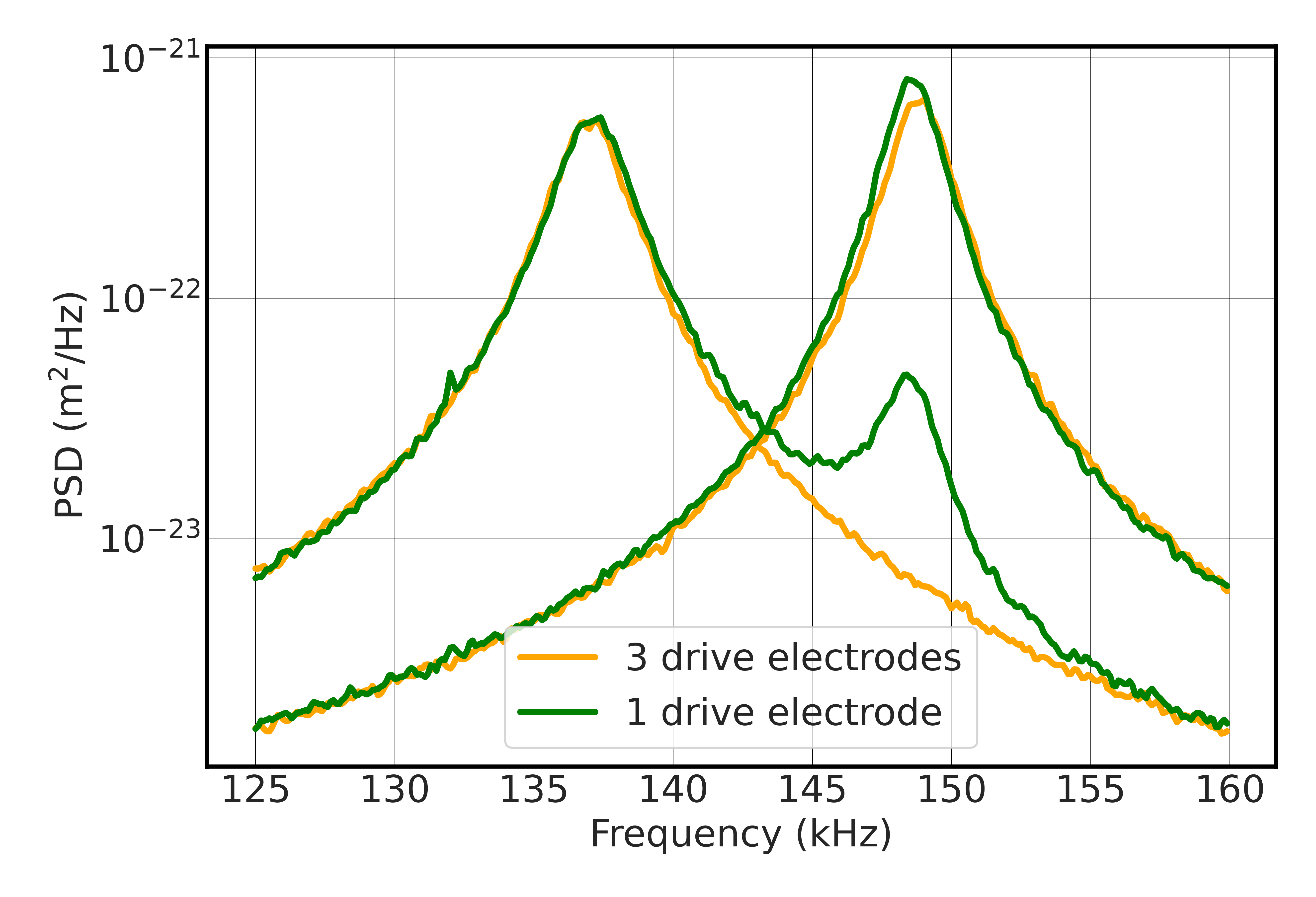}
\centering
\caption{\label{fig:figureVD2}The PSD for the nanoparticle cooled via the two different electrode configurations. The PSD have similar linewidth and height, indicating cooling to comparable temperatures. A significant increase in the height of the signal for the $y$ motion in the $x$ detection indicates significant cross-talk.}
\end{figure}

The experimental power spectral densities (PSDs) can be seen in Fig.~\ref{fig:figureVD2}. There is a significant increase in the height of the $y$ motion appearing in the $x$ detection, when cooled using the one drive electrode, indicating cross-talk has been induced. From the three drive electrodes, there is no change in height and no indication of cross-talk.

\begin{figure}
\includegraphics[width=1\columnwidth]{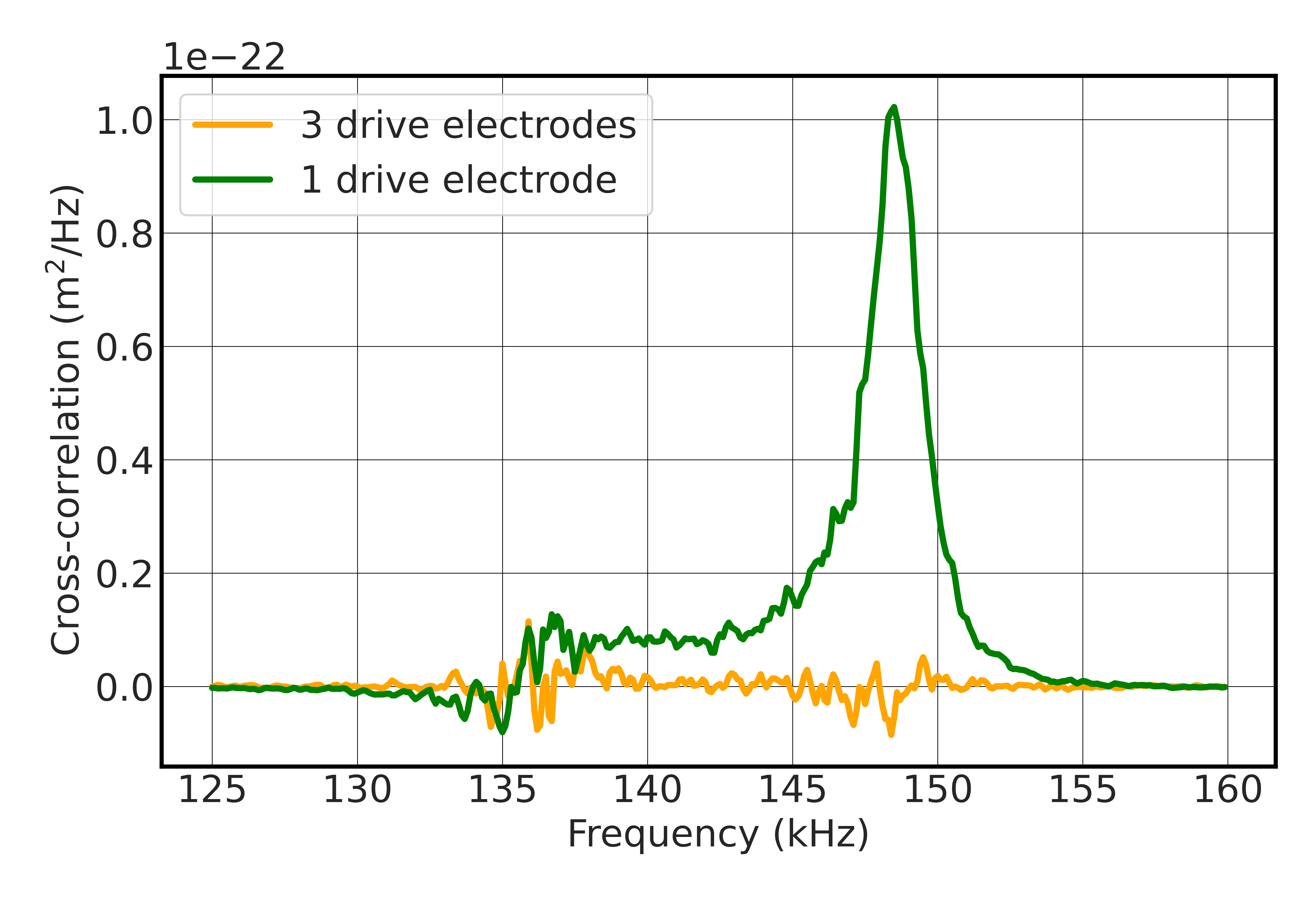}
\centering
\caption{\label{fig:figureVD3} The cross correlation mechanical spectra for the nanoparticle cooled via the two different electrode configurations. The height of the cross-correlation indicates the amount of correlations induced at that frequency. Minimal correlations are induced for the 3 drive electrode case compared to the 1 drive electrode configuration. The cross-correlation when no feedback is applied is subtracted from
the spectra to remove correlations due to detector misalignment.}
\end{figure}

To further illustrate the effect of the cross-talk, the mechanical cross-correlation spectra $S_{xy}(\omega)=x(\omega) \ast y(\omega)$  is considered \cite{Gosling2024Sensing}. The cross-correlation indicates how correlated the $x$ and $y$ directions are. The cross-correlation spectra can be seen in Fig.~\ref{fig:figureVD3}.  It is expected that the cross-correlation should be 0 when the motion is completely uncoupled with no cross talk. The magnitude gives the strength of correlation and the sign whether it is correlated or anti-correlated. From Fig.~\ref{fig:figureVD3}, there are clear correlations induced by the one drive electrode configuration at the $y$ frequency, with minimal correlations in the feedback when three drive electrodes were used. This is in agreement with the PSD in Fig.~\ref{fig:figureVD2}. The four electrode configuration has reduced the cross-talk by an order of magnitude.

\begin{figure}[h!]
\includegraphics[width=1\columnwidth]{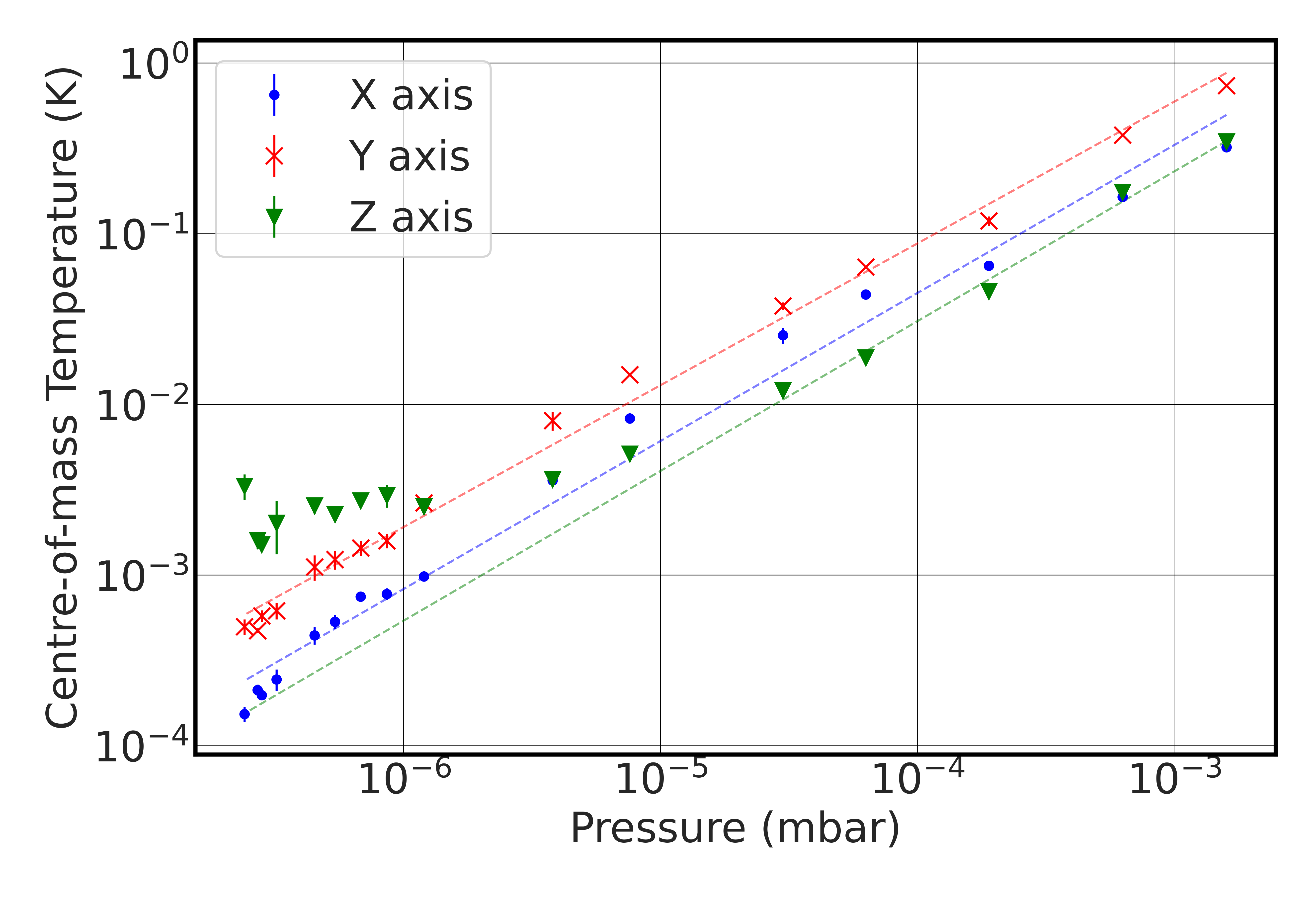}
\centering
\caption{\label{fig:figTvP}The centre-of-mass temperature as a function of pressure for the three different degrees of freedom. The lowest temperature reached was $0.15\pm 0.02$  mK, $0.47\pm0.02$ mK, and $1.5\pm0.1$ mK for $x,y$, and $z$ respectively.}
\end{figure}

\emph{Cooling a particle in 3D with reduced crosstalk} -To test the capabilities of the cooling method, the relationship between cooling and pressure was investigated. It is expected that the centre-of-mass temperature should decrease linearly with pressure, as the thermal gas damping is removed \cite{gieseler2012subkelvin}. The centre-of-mass temperatures at various pressures is shown in Fig.~\ref{fig:figTvP}.

As can be seen in Fig.~\ref{fig:figTvP}, the linear relationship holds well for the $x$ and $y$ axes in the pressure regime used. For these axes, there are deviations away from the straight line fit. These are most likely due to the uncertainty in the pressure reading from the type of gauge used. However, the linear relationship breaks down for the $z$ axis below $10^{-6}$ mbar, as the temperature plateaus with decreasing pressure. This indicates that there is an additional heating mechanism other than the gas which is preventing the temperature from being reduced. The likely cause of this heating is the poor detection efficiency of the $z$ motion. The imprecision noise for the $x$ axis is approximately $1 \times 10^{-24} \textup{m}^2/\textup{Hz}$, approximately $4 \times 10^{-23} \textup{m}^2/\textup{Hz}$ for the $z$ axis and approximately $2 \times 10^{-24} \textup{m}^2/\textup{Hz}$ for the $y$ axis.

At the lowest possible pressure of $2.4 \times 10^{-7}$ mbar, a temperature of  $0.15\pm 0.02$ mK was reached for the $x$ axis. For the $y$ axis, the lowest temperature of $0.47\pm0.02$ mK was achieved at  $2.7 \times 10^{-7}$ mbar and for $z$ the lowest temperature of $1.5\pm0.1$ mK was achieved at $2.8 \times 10^{-7}$ mbar. The difference in pressure for the lowest temperatures was mostly likely due to fluctuations in laser power and the feedback signal during the measurement.  By considering $\left \langle n \right \rangle =\frac{1}{e^{\frac{\hbar \omega_0}{k_B T_{CM}}}-1}$ the phonon occupancy of each oscillation mode can be calculated. For the lowest achieved temperature, this corresponds to an occupation number of $24\pm 2$, $65\pm 2$, and $950\pm70$ for $x,~y$ and $z$ respectively.

As discussed in Eq.~\ref{eq:6.3}, another source of noise is the voltage noise, $n_{el}$, associated with the electronics used. For this experiment, the dominant source of this noise is from the high voltage amplifiers used, with a total voltage noise, $S_{el}\approx10^{-9}~\textup{V}^2/\textup{Hz}$. Whilst the gas collisions are the limiting factor for the centre-of-mass temperatures at this pressure, the voltage noise might contribute at lower pressures. Indeed, this additional noise heats the nanoparticle's motion and can become a particular issue when the nanoparticle is highly charged. The total feedback force acting on a nanoparticle, without the presence of voltage noise, is given by the product of the nanoparticle's charge and the feedback gain. For a highly charged nanoparticle, less feedback gain is required to achieve the same level of cooling. However, this would increase the effect of voltage noise which translates to tighter requirements in terms of maximum voltage noise of the electronics used

Using this feedback technique, the nanoparticle has been cooled from room temperature to tens of phonons in two of the three directions.  At this phonon occupancy, quantum mechanical effects can start to be investigated \cite{tebbenjohanns2020motional}. To reach lower phonon occupancies, detection efficiency in these two directions would need to be improved \cite{tebbenjohanns2019optimal,dinter2024three}.

\emph{Conclusion} - We have described a method for implementing electrical cold damping of an optically levitated particle which enables significantly reduced cross-talk between the mechanical modes. This is achieved through the use of a four electrode configuration which allows cancellation of the feedback electrical fields at the nanoparticle, ensuring that the feedback acts along only the intended direction. This electrode configuration allows for applications of an electric field with arbitrary directional components with the potential for extension to cooling of librational motion \cite{Blakemore2022Librational}. 

One of the advantages of levitated optomechanics is the geometric interpretation of the oscillatory modes means that these degrees of freedom can be uncoupled from each other to first order in position.  Our implementation of cooling with reduced cross-talk will be important when used for directional force sensing in 3D where we require that the oscillatory modes to be uncorrelated \cite{Gosling2024Sensing}. This scheme could also be important in thermodynamic experiments when the heating rates of individual modes are to be considered \cite{Rahman2023Measurement,Jain2016Direct}. This scheme may find application in optomechncial dark matter searches where directional forces are expected \cite{Kilian2024Dark,ahlen2010case}. Although we have applied this to single beam levitation in an optical tweezers the principles behind this method are general and can be achieved with other configurations of electrode design and can also be extended to electrical feedback cooling in other levitation platforms including Paul traps~\cite{penny2021performance}.

\section*{Acknowledgement}
J.M.H.G., M.R., A.P., and P.F.B. acknowledge funding from the
EPSRC via Grant Nos. EP/S000267/1 and EP/W029626/1. 
J.M.H.G., M.R., A.P. and P.F.B. acknowledge funding from the
H2020-EU.1.2.1 TEQ Project through Grant Agreement ID 766900.  J.M.H.G. acknowledges funding from the Science and Technology Facilities Council (STFC) Grant No.
ST/W006170/1.

\bibliographystyle{nicebib}
\bibliography{feedback}

\end{document}